\documentclass[aps,prd,twocolumn,showpacs,superscriptaddress,nofootinbib,preprintnumbers]{revtex4-1}
\pdfoutput=1
\usepackage{amssymb,amsmath,latexsym,mathrsfs}
\usepackage{graphicx,subfigure}
\usepackage{epsfig}
\usepackage{varioref,xr-hyper}
\usepackage{color}
\usepackage{multirow}
\usepackage{tikz}
\usepackage{array}
\usepackage{hyperref}
\usepackage{wasysym}
\usepackage{color}
\usepackage{float}
\usepackage[utf8]{inputenc}
\usepackage[T1]{fontenc}

\makeatletter
\def\thickhline{%
  \noalign{\ifnum0=`}\fi\hrule \@height \thickarrayrulewidth \futurelet
   \reserved@a\@xthickhline}
\def\@xthickhline{\ifx\reserved@a\thickhline
               \vskip\doublerulesep
               \vskip-\thickarrayrulewidth
             \fi
      \ifnum0=`{\fi}}
\makeatother

\newlength{\thickarrayrulewidth}
\begin{document}
\title{Non-minimal dark sector physics and cosmological tensions}

\author{Eleonora Di Valentino}
\email{eleonora.divalentino@manchester.ac.uk}
\affiliation{Jodrell Bank Center for Astrophysics, School of Physics and Astronomy, University of Manchester, Oxford Road, Manchester M13 9PL, United Kingdom}

\author{Alessandro Melchiorri}
\email{alessandro.melchiorri@roma1.infn.it}
\affiliation{Department of Physics and Istituto Nazionale di Fisica Nucleare (INFN), University of Rome ``La Sapienza'', Piazzale Aldo Moro 2, 00185 Rome, Italy}

\author{Olga Mena}
\email{omena@ific.uv.es}
\affiliation{Instituto de F\'{i}sica Corpuscular (IFIC), University of Valencia-CSIC, E-46980, Valencia, Spain}

\author{Sunny Vagnozzi}
\email{sunny.vagnozzi@ast.cam.ac.uk}
\affiliation{Kavli Institute for Cosmology (KICC) and Institute of Astronomy, University of Cambridge, Madingley Road, Cambridge CB3 0HA, United Kingdom}

\date{\today}

\begin{abstract}
We explore whether non-standard dark sector physics might be required to solve the existing cosmological tensions. The properties we consider in combination are: \textit{(a)} an interaction between the dark matter and dark energy components, and \textit{(b)} a dark energy equation of state $w$ different from that of the canonical cosmological constant $w=-1$. In principle, these two parameters are independent. In practice, to avoid early-time, superhorizon instabilities, their allowed parameter spaces are correlated. Moreover, a clear degeneracy exists between these two parameters in the case of Cosmic Microwave Background (CMB) data. We analyze three classes of extended interacting dark energy models in light of the 2019 \textit{Planck} CMB results and Cepheid-calibrated local distance ladder $H_0$ measurements of Riess et al. (R19), as well as recent Baryon Acoustic Oscillation (BAO) and Type Ia Supernovae (SNeIa) distance data. We find that in \textit{quintessence} coupled dark energy models, where $w > -1$, the evidence for a non-zero coupling between the two dark sectors can surpass the $5\sigma$ significance. Moreover, for both Planck+BAO or Planck+SNeIa, we found a preference for $w>-1$ at about three standard deviations. Quintessence models are, therefore, in excellent agreement with current data when an interaction is considered. On the other hand, in \textit{phantom} coupled dark energy models, there is no such preference for a non-zero dark sector coupling. All the models we consider significantly raise the value of the Hubble constant easing the $H_0$ tension. In the interacting scenario, the disagreement between Planck+BAO and R19 is considerably reduced from $4.3\sigma$ in the case of $\Lambda$CDM to about $2.5\sigma$. The addition of low-redshift BAO and SNeIa measurements leaves, therefore, some residual tension with R19 but at a level that could be justified by a statistical fluctuation. Bayesian evidence considerations mildly disfavour both the coupled quintessence and phantom models, while mildly favouring a coupled vacuum scenario, even when late-time datasets are considered. We conclude that non-minimal dark energy cosmologies, such as coupled quintessence, phantom, or vacuum models, are still an interesting route towards softening existing cosmological tensions, even when low-redshift datasets and Bayesian evidence considerations are taken into account.
\end{abstract}


\maketitle

\section{Introduction}
\label{sec:intro}

The canonical $\Lambda$CDM scenario has proven to provide an excellent match to observations at high and low redshift, see for instance~\cite{Riess:1998cb,Perlmutter:1998np,Dunkley:2010ge,Hinshaw:2012aka,Ade:2013zuv,
Story:2014hni,Ade:2015xua,Alam:2016hwk,Troxel:2017xyo,Aghanim:2018eyx}. Despite its enormous success, there are some tensions among the values of cosmological parameters inferred from independent datasets~\cite{Freedman:2017yms,DiValentino:2017gzb,DiValentino:2018gcu}. The most famous and persisting one is that related to the value of the Hubble constant $H_0$ as measured  from \textit{Planck} Cosmic Microwave Background (CMB) data ($h = (0.6737 \pm 0.0054)$~\cite{Aghanim:2018eyx}) versus the value extracted from Cepheid-calibrated local distance ladder measurements (\textit{R19}, $h=(0.7403 \pm 0.0142)$~\cite{Riess:2019cxk}), referred to as the \textit{$H_0$ tension}, with $h=H_0/(100\,{\rm km}\,{\rm s}^{-1}\,{\rm Mpc}^{-1})$~\footnote{In Ref.~\cite{Jackson:2007ug,Verde:2019ivm} the reader can find complete reviews comparing the CMB and local determinations of $H_0$.}. This tension now reaches the $4.4\sigma$ level. 

Two main avenues have been followed to solve the $H_0$ tension. The first one is based on the possibility that \textit{Planck} and/or the local distance ladder measurement of $H_0$ suffer from unaccounted systematics~\footnote{See e.g.~\cite{Spergel:2013rxa,Addison:2015wyg,Aghanim:2016sns,
Lattanzi:2016dzq,Huang:2018xle} for studies of possible systematics in the context of \textit{Planck} and e.g.~\cite{Rigault:2013gux,Rigault:2014kaa,Scolnic:2017caz,
Jones:2018vbn,Rigault:2018ffm} in the context of the local distance ladder measurement. Local measurements other than the R19 one exist, but most of them appear to consistently point towards values of $H_0$ significantly higher than the CMB one (see e.g.~\cite{Efstathiou:2013via,Cardona:2016ems,Feeney:2017sgx,Dhawan:2017ywl,
Follin:2017ljs,Gomez-Valent:2018hwc,Birrer:2018vtm,Burns:2018ggj,Jimenez:2019onw,
Collett:2019hrr,Wong:2019kwg,Freedman:2019jwv,Liao:2019qoc,Reid:2019tiq,Jee:2019hah}).}. The second more intriguing possibility is that the $H_0$ tension might be the first sign for physics beyond the concordance $\Lambda$CDM model. The most economical possibilities in this direction involve phantom dark energy (\textit{i.e.} a dark energy component with equation of state $w<-1$) or some form of dark radiation (so as to raise $N_{\rm eff}$ beyond its canonical value of $3.046$)~\cite{DiValentino:2016hlg,Bernal:2016gxb,Vagnozzi:2019ezj}. However, in recent years, a number of other exotic scenarios attempting to address the $H_0$ tension have been examined, including (but not limited to) decaying dark matter (DM), interactions between DM and dark radiation, a small spatial curvature, an early component of dark energy (DE), and modifications to gravity (see e.g.~\cite{Alam:2016wpf,Qing-Guo:2016ykt,Ko:2016uft,Karwal:2016vyq,Chacko:2016kgg,
Zhao:2017cud,Vagnozzi:2017ovm,Agrawal:2017rvu,Benetti:2017gvm,Feng:2017nss,Zhao:2017urm,
DiValentino:2017zyq,Gariazzo:2017pzb,Dirian:2017pwp,Feng:2017mfs,Renk:2017rzu,
Yang:2017alx,Buen-Abad:2017gxg,Raveri:2017jto,DiValentino:2017rcr,DiValentino:2017oaw,
Khosravi:2017hfi,Peirone:2017vcq,Benetti:2017juy,Mortsell:2018mfj,Vagnozzi:2018jhn,
Nunes:2018xbm,Poulin:2018zxs,Kumar:2018yhh,Banihashemi:2018oxo,DEramo:2018vss,
Guo:2018ans,Graef:2018fzu,Yang:2018qmz,Banihashemi:2018has,Aylor:2018drw,
Poulin:2018cxd,Kreisch:2019yzn,Pandey:2019plg,Vattis:2019efj,Colgain:2019pck,
Agrawal:2019lmo,Li:2019san,Yang:2019jwn,Colgain:2019joh,Keeley:2019esp,Li:2019yem,
DiValentino:2019exe,Archidiacono:2019wdp,Desmond:2019ygn,Yang:2019nhz,Nesseris:2019fwr,
Visinelli:2019qqu,Cai:2019bdh,Schoneberg:2019wmt,Pan:2019hac,DiValentino:2019dzu,
Xiao:2019ccl,Panpanich:2019fxq,Knox:2019rjx,Ghosh:2019tab,Escudero:2019gvw,Yan:2019gbw,
Banerjee:2019kgu,Yang:2019uog,Cheng:2019bkh,Sakstein:2019fmf,Liu:2019awo,
Anchordoqui:2019amx,Wang:2019isw,Mazo:2019pzn,Pan:2020zza,Yang:2020zuk,
Lyu:2020lwm,Yang:2020uga} for an incomplete list of recent papers).~\footnote{Other scenarios worth mentioning include the possibility that properly accounting for cosmic variance (due to the fact that a limited sample of the Hubble flow is observed) enlarges the uncertainty of the locally determined $H_0$ to the point that the tension is alleviated~\cite{Marra:2013rba,Wojtak:2013gda,Wu:2017fpr,Camarena:2018nbr,
Bengaly:2018xko}, or that local measurements might be biased by the presence of a local void~\cite{Keenan:2013mfa,Romano:2016utn,Fleury:2016fda,Hoscheit:2018nfl,
Shanks:2018rka} (see however e.g.~\cite{Odderskov:2014hqa,Kenworthy:2019qwq} for criticisms on both these possibilities). From the theoretical side models of running vacuum, motivated by QFT corrections in curved spacetime, are instead among the most theoretically well-motivated solutions to the $H_0$ tension (see for example~\cite{Sola:2017znb,Gomez-Valent:2018nib,Rezaei:2019xwo,Sola:2019jek}).}
 
From the theoretical perspective, interactions between DM and DE beyond the purely gravitational ones are not forbidden by any fundamental symmetry in nature~\cite{Amendola:2007yx,Micheletti:2009pk,Pavan:2011xn,Bolotin:2013jpa,
Costa:2014pba,Ludwick:2019yso} and could help addressing the so called coincidence or \textit{why now?} problem~\cite{Hu:2006ar,Sadjadi:2006qp,delCampo:2008jx,Dutta:2017kch,Dutta:2017fjw} , see e.g.~\cite{Farrar:2003uw,Barrow:2006hia,Amendola:2006dg,He:2008tn,Valiviita:2008iv,
Gavela:2009cy,CalderaCabral:2009ja,Majerotto:2009np,Abdalla:2009mt,Honorez:2010rr,
Clemson:2011an,Pan:2012ki,Salvatelli:2013wra,Yang:2014vza,Yang:2014gza,Nunes:2014qoa,
Faraoni:2014vra,Pan:2014afa,Ferreira:2014cla,Tamanini:2015iia,Li:2015vla,Murgia:2016ccp,
Nunes:2016dlj,Yang:2016evp,Pan:2016ngu,Sharov:2017iue,An:2017kqu,Santos:2017bqm,
Mifsud:2017fsy,Kumar:2017bpv,Guo:2017deu,Pan:2017ent,An:2017crg,Costa:2018aoy,
Wang:2018azy,vonMarttens:2018iav,Yang:2018qec,Martinelli:2019dau,Li:2019loh,
Yang:2019vni,Bachega:2019fki,Yang:2019uzo,Li:2019ajo,
Mukhopadhyay:2019jla,Carneiro:2019rly,Kase:2019veo,Yamanaka:2019aeq,Yamanaka:2019yek} and Ref.~\cite{Wang:2016lxa} for a recent comprehensive review on interacting dark sector models, motivated by the idea of coupled quintessence~\cite{Wetterich:1994bg,Amendola:1999dr,Amendola:1999er,
Mangano:2002gg,Zhang:2005rg,Saridakis:2010mf,Barros:2018efl,
DAmico:2018mnx,Liu:2019ygl}.~\footnote{See also~\cite{Benisty:2017eqh,Benisty:2018qed,Benisty:2018oyy,
Anagnostopoulos:2019myt,Benisty:2019jqz} for examples of models of unified interacting DM-DE.} These models may also be an interesting key towards solving some existing cosmological tensions~\cite{Salvatelli:2014zta,Kumar:2016zpg,Xia:2016vnp,Kumar:2017dnp,Yang:2017ccc,
Feng:2017usu,Yang:2018ubt,Yang:2018xlt,Yang:2018uae,
Li:2018ydj,Kumar:2019wfs,Pan:2019jqh,DiValentino:2017iww,Yang:2017ccc,Feng:2017usu,
Yang:2018ubt,Yang:2018xlt,Yang:2018uae,Li:2018ydj,Kumar:2019wfs,Pan:2019jqh,
Yang:2019uzo,Pan:2019gop,Benetti:2019lxu}.

We have recently shown that one particular and well-studied interacting DE model is still a viable solution to the $H_0$ tension in light of the 2019 \textit{Planck} CMB and local measurement of $H_0$~\cite{DiValentino:2019ffd}. However, our study in~\cite{DiValentino:2019ffd} considered a minimal dark energy scenario, where the interacting DE component is essentially a cosmological constant (see~\cite{Bamba:2012cp} for a recent review on dark energy models). In this work, we allow for more freedom in the DE sector, considering a more generic DE component with an equation of state $w$ not necessarily equal to $-1$. We here study in more detail the properties of DE required to solve the $H_0$ tension, analyzing the suitable values of the coupling ($\xi$) and the equation of state ($w$) for the DE component which can ameliorate the Hubble tension.  While these two parameters are, in principle, independent, the potential presence of early-time superhorizon instabilities results in their viable parameter spaces being correlated.

The rest of this paper is then organized as follows. Section~\ref{sec:interacting} reviews the basic equations governing the cosmology of extended interacting dark energy models, briefly discussing their stability and initial conditions. The methodology and datasets adopted in our numerical studies are presented in Sec.~\ref{sec:data}, whereas in Sec.~\ref{sec:results} we present our results. We conclude in Sec.~\ref{sec:conclusions}.

\section{Extended interacting dark energy models}
\label{sec:interacting}

Interacting dark energy models (IDE in what follows) are characterized by a modification to the usual conservation equations of the DM and DE energy-momentum tensors $T^{\mu\nu}_c$ and $T^{\mu\nu}_x$ (which would usually read $\nabla_{\nu}T^{\mu\nu}_c=\nabla_{\nu}T^{\mu\nu}_x=0$), which now read~\cite{Valiviita:2008iv,Gavela:2009cy}:
\begin{eqnarray}
\nabla_{\nu}T^{\mu\nu}_c &=& \frac{Qu^{\mu}}{a}\,,\\
\nabla_{\nu}T^{\mu\nu}_x &=& -\frac{Qu^{\mu}}{a}\,,
\label{eq:continuity}
\end{eqnarray}
where $a$ is the scale factor and the DM-DE interaction rate is given by $Q$:
\begin{eqnarray}
Q = \xi{\cal H}\rho_x\,,
\label{eq:coupling}
\end{eqnarray}
with $\xi$ is a dimensionless number quantifying the strength of the DM-DE coupling. From now on, we shall refer to $\xi$ as the DM-DE coupling.  Notice that $Q>0$ and $Q<0$ indicate, respectively, energy transfer from DE to DM and viceversa, or a possible decay of DE into DM and viceversa, depending on the details of the underlying model.

At the background level, for a pressureless cold DM component and a DE component with equation of state (EoS) $w$,  the evolution of the background DM and DE energy densities are~\cite{Gavela:2009cy}:
\begin{eqnarray}
\label{eq:rhoc}
\rho_c &=& \frac{\rho^0_c}{a^3}+\frac{\rho^0_x}{a^3} \left [ \frac{\xi}{3w+\xi} \left ( 1-a^{-3w-\xi} \right ) \right ]\,, \\
\label{eq:rhox}
\rho_x &=& \frac{\rho^0_x}{a^{3(1+w)+\xi}} \,,
\end{eqnarray}
where $\rho^0_c$ and $\rho^0_x$ are the DM and DE energy densities today, respectively. At the linear perturbation level, and setting the DE speed of sound $c_{s,x}^2=1$, the evolution of the DM and DE density perturbations ($\delta_c$, $\delta_x$) and velocities ($\theta_c$, $\theta_x$) are given by:
\small
\begin{eqnarray}
\label{eq:deltac}
\dot{\delta}_c &=& -\theta_c - \frac{1}{2}\dot{h} +\xi{\cal H}\frac{\rho_x}{\rho_c}(\delta_x-\delta_c)+\xi\frac{\rho_x}{\rho_c} \left ( \frac{kv_T}
{3}+\frac{\dot{h}}{6} \right )\,, \\
\label{eq:thetac}
\dot{\theta}_c &=& -{\cal H}\theta_c\,,\\
\label{eq:deltax}
\dot{\delta}_x &=& -(1+w) \left ( \theta_x+\frac{\dot{h}}{2} \right )-\xi \left ( \frac{kv_T}{3}+\frac{\dot{h}}{6} \right ) \nonumber \\
&&-3{\cal H}(1-w) \left [ \delta_x+\frac{{\cal H}\theta_x}{k^2} \left (3(1+w)+\xi \right ) \right ]\,,\\
\label{eq:thetax}
\dot{\theta}_x &=& 2{\cal H}\theta_x+\frac{k^2}{1+w}\delta_x+2{\cal H}\frac{\xi}{1+w}\theta_x-\xi{\cal H}\frac{\theta_c}{1+w}\,,
\end{eqnarray}
\normalsize
where $h$ is the usual synchronous gauge metric perturbation. In addition, $v_T$ is the center of mass velocity for the total fluid, whose presence is required by gauge invariance considerations~\cite{Gavela:2010tm}:
\begin{eqnarray}
v_T = \frac{\sum_i \rho_i q_i}{\sum_i \left ( \rho_i + P_i \right )}\,,
\label{eq:vt}
\end{eqnarray}
where the index $i$ runs over the various species (whose energy densities and pressures are $\rho_i$ and $P_i$), and $q_i$ is the heat flux of species $i$, given by:
\begin{eqnarray}
q_i = \frac{ \left ( \rho_i + P_i \right ) \theta_i}{kP_i}\,.
\label{eq:qi}
\end{eqnarray}
The initial conditions for the DE perturbations $\delta_x$ and $\theta_x$ also need to be modified to the following~\cite{Gavela:2010tm}:
\begin{eqnarray}
\delta_x^{\rm in}(\eta) &=& \frac{1+w+\xi/3}{12w^2-2w-3w\xi+7\xi-14}\delta_{\gamma}^{\rm in}(\eta)\nonumber \\
& \times & \frac{3}{2}(2\xi-1-w)\,, \\
\theta_x^{\rm in}(x) &=& \frac{3}{2}\frac{\eta(1+w+\xi/3)}{2w+3w\xi+14-12w^2-7\xi}\delta_{\gamma}^{\rm in}(\eta)\,,
\label{eq:initialconditions}
\end{eqnarray}
where $\eta= k \tau$.

Finally, besides affecting the evolution of the background and the perturbation evolution, as well as requiring suitable initial conditions, the presence of a DM-DE coupling may affect the stability of the interacting system. Apart from the gravitational instabilities present when $w=-1$~\cite{Valiviita:2008iv,He:2008si}, there may also be early-time instabilities~\cite{Valiviita:2008iv,He:2008si,Jackson:2009mz,Gavela:2009cy,Gavela:2010tm,Clemson:2011an}, and avoiding them leads to imposing stability conditions on $w$ and $\xi$. Therefore, within the model in question, even though in principle the two parameters $\xi$ and $w$ describing the dark energy physics sector are independent, it turns out that only two distinct classes of models remain possible: essentially, the signs of $\xi$ and $1+w$ have to be opposite. In one class of models $\xi>0$ and $w<-1$ (and thus energy flows from DE to DM), and in the second one $\xi<0$ and $w>-1$ (thus energy transfer occurs from DM to DE).~\footnote{Other possibilities considered in the literature to address these two types of instabilities include an extension of the parametrized post-Friedmann approach to the IDE case~\cite{Li:2014eha,Li:2014cee,Guo:2017hea,Zhang:2017ize,Guo:2018gyo,Dai:2019vif}, as well as considering phenomenological coupling functions $Q$ depending on the DE EoS $w$~\cite{Yang:2017zjs,Yang:2017ccc,Yang:2018euj,Yang:2018xlt}.} Also, as it is clear from Eq.~(\ref{eq:rhoc}), even when the aforementioned instability-free prescriptions are considered, one needs to ensure that the DM energy density remains positive by requiring $\xi<-3w$. This is not a problem when $\xi<0$ and $w>-1$, since accelerated expansion requires $w<-1/3$, and therefore $w$ cannot take positive values, meaning that $\xi<0$ automatically implies $\xi<-3w$. For the $\xi>0$ and $w<-1$ case, the condition $\xi<-3w$ is not automatically satisfied, and it needs to be imposed as an extra constraint on the allowed parameter spaces.

\begin{figure*}[th]
\begin{center}
	\includegraphics[width=0.49\textwidth]{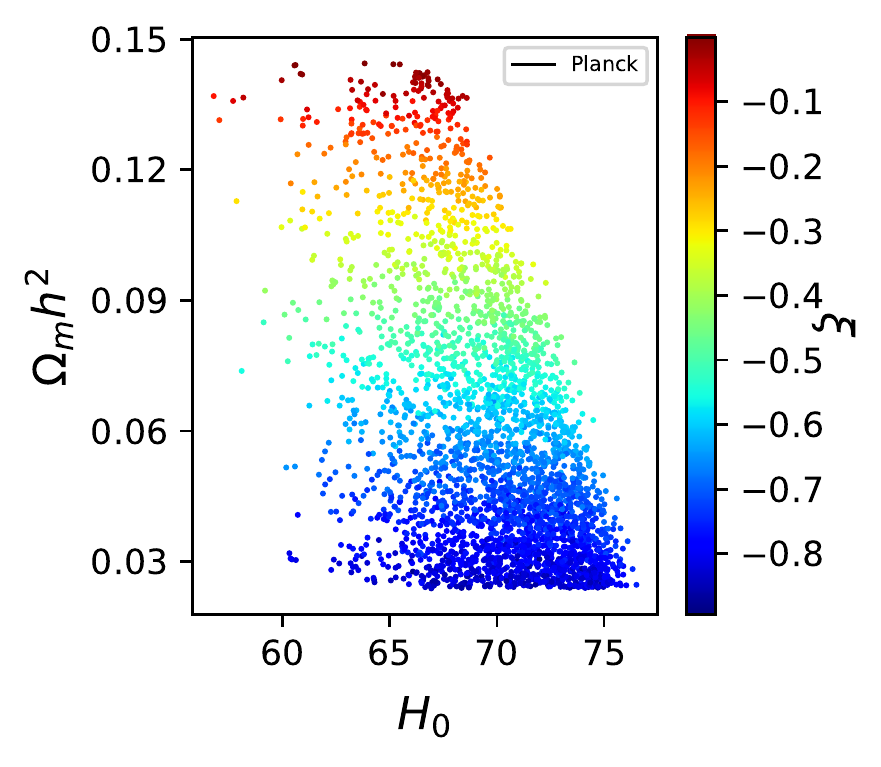}
	\includegraphics[width=0.49\textwidth]{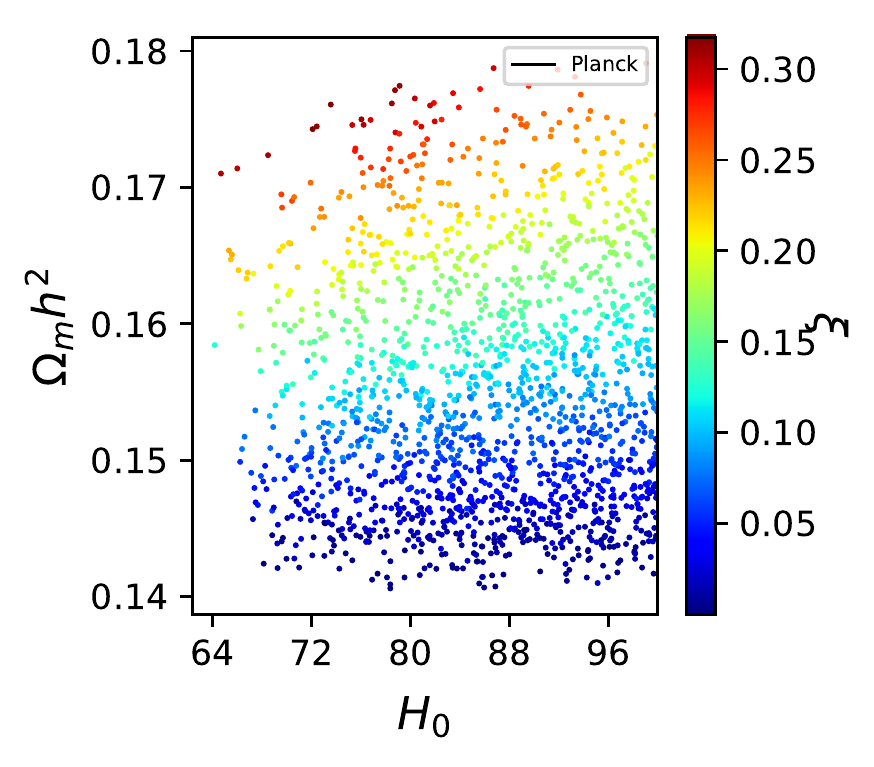}
	\caption{Left (right) panel: Samples from Planck chains in the ($H_0$, $\Omega_m h^2$) plane for the $\xi q$CDM ($\xi p$CDM) model, color-coded by $\xi$.}
	\label{fig:tri}
	\end{center}
\end{figure*}

\section{Models and datasets}
\label{sec:data}
The parameter space of the IDE model we consider is described by the usual six cosmological parameters of $\Lambda$CDM, complemented by one or two additional parameters depending on whether we allow the dark energy equation of state $w$ to vary freely. We recall that the six parameters of the $\Lambda$CDM model are the baryon and cold DM physical density parameters $\Omega_bh^2$ and $\Omega_ch^2$, the angular size of the sound horizon at decoupling $\theta_s$ (given by the ratio between the sound horizon to the angular diameter distance at decoupling), the optical depth to reionization $\tau$, and the amplitude and tilt of the primordial power spectrum of scalar fluctuations $A_s$ and $n_s$. To these $6$ cosmological parameters, we add the DM-DE coupling $\xi$ and the DE EoS $w$.

The stability issue discussed in Sec.~\ref{sec:interacting} will influence the choice of priors on the cosmological parameters. Ideally, we would want to consider two types of cosmological models: $\Lambda$CDM+$\xi$ (seven parameters) and $\Lambda$CDM+$\xi$+$w$ (eight parameters). Technically speaking, within the baseline $\Lambda$CDM model, the DE EoS would be fixed to $w=-1$. However, as we discussed in Sec.~\ref{sec:interacting}, in the case of IDE models, this leads to gravitational instabilities, which undermine the viability of the model. Therefore, na\"{i}vely considering a baseline $\Lambda$CDM+$\xi$ model would not work and we fix the DE EoS to $w=-0.999$ instead, an approach already adopted in~\cite{Salvatelli:2013wra,DiValentino:2019ffd}. Indeed, for $\Delta w \equiv 1+w$ sufficiently small, Eqs.~(\ref{eq:deltax},\ref{eq:thetax}) are essentially only capturing the effect of the DM-DE coupling $\xi$, while at the same time the absence of gravitational instabilities is guaranteed. To avoid early-time instabilities, we also require $\xi<0$. We refer to this model as $\xi\Lambda$CDM or coupled vacuum scenario.

We then extend the baseline coupled vacuum $\xi\Lambda$CDM model by allowing the DE EoS $w$ to vary. To satisfy the stability conditions, see Sec.~\ref{sec:interacting}, we consider two different cases: one where $\xi>0$ and $w<-1$ (which we refer to as $\xi p$CDM model, where the ``\textit{p}'' reflects the fact that the DE EoS lies in the phantom regime), and one where $\xi<0$ and $w>-1$ (which we refer to as $\xi q$CDM model, where the ``\textit{q}'' reflects the fact that the DE EoS lies in the quintessence regime).~\footnote{See e.g.~\cite{Shahalam:2015sja,Shahalam:2017fqt} for concrete examples of construction and dynamical system analyses of coupled quintessence and coupled phantom models.} The three interacting dark energy models we consider in this work, and in particular the values of $w$ and $\xi$ allowed by stability conditions therein, are summarized in Tab.~\ref{tab:models}.
\begin{table}[!b]
\begin{tabular}{|c||c|c|c|}
\hline
\textbf{Model} & DE EoS & DM-DE coupling & Energy flow \\
\hline
\hline
$\xi\Lambda$CDM & $w=-0.999$ & $\xi<0$ & DM$\to$DE \\
\hline
$\xi p$CDM & $w<-1$ & $\xi>0\,,\quad \xi<-3w$ & DE$\to$DM  \\
\hline
$\xi q$CDM & $w>-1$ & $\xi<0$ & DM$\to$DE \\
\hline
\end{tabular}
\caption{Summary of the three interacting dark energy models considered in this work. For all three cases, we report the values allowed for the DE EoS $w$ and the DM-DE coupling $\xi$ ensuring that gravitational instabilities, early-time instabilities, and unphysical values for the DM energy density are avoided, as well as the direction of energy flow (DE$\to$DM or DM$\to$DE). For all models, we vary the six usual parameters of the $\Lambda$CDM model.}
\label{tab:models}
\end{table}

Having described the three models we consider in this work, we now proceed to describe the datasets we adopt. We first consider measurements of CMB temperature and polarization anisotropies, as well as their cross-correlations. This dataset is called Planck TT,TE,EE+lowE in~\cite{Aghanim:2018eyx}, whereas we refer to it as \textit{Planck}. We then include the lensing reconstruction power spectrum obtained from the CMB trispectrum analysis~\cite{Aghanim:2018oex}, which we refer to as \textit{lensing}.

In addition to CMB data, we also consider Baryon Acoustic Oscillation (BAO) measurements from the 6dFGS~\cite{Jones:2009yz,Beutler:2011hx}, SDSS-MGS~\cite{York:2000gk,Ross:2014qpa}, and BOSS DR12~\cite{Alam:2016hwk} surveys, and we shall refer to the combination of these BAO measurements as \textit{BAO}.  Supernovae Type Ia (SNeIa) distance moduli data from the \textit{Pantheon} sample~\cite{Scolnic:2017caz}, the largest spectroscopically confirmed SNeIa sample consistent of distance moduli for 1048 SNeIa, are also included in our numerical analyses, and we refer to this dataset as \textit{Pantheon}. We also consider a Gaussian prior on the Hubble constant $H_0=74.03\pm1.42$ km/s/Mpc, as measured by the SH0ES collaboration in~\cite{Riess:2019cxk}, and we refer to it  as \textit{R19}. 

Finally, we consider a case where we combine all the aforementioned datasets (\textit{Planck}, \textit{lensing}, \textit{BAO}, \textit{Pantheon}, and \textit{R19}). We refer to this dataset combination as \textit{All19}.

We modify the Boltzmann solver \texttt{CAMB}~\cite{Lewis:1999bs} to incorporate the effect of the DM-DE coupling as in Eqs.~(\ref{eq:deltac}-\ref{eq:thetax}). We sample the posterior distribution of the cosmological parameters by making use of Markov Chain Monte Carlo (MCMC) methods, through a suitably modified version of the publicly available MCMC sampler \texttt{CosmoMC}~\cite{Lewis:2002ah,Lewis:2013hha}. We monitor the convergence of the generated MCMC chains through the Gelman-Rubin parameter $R-1$~\cite{Gelman:1992zz}, requiring $R-1<0.01$ for our MCMC chains to be considered converged.

In addition to performing parameter estimation, we also perform a model comparison analysis. In particular, we use our MCMC chains to compute the Bayesian evidence for the three interacting dark energy models ($\xi\Lambda$CDM, $\xi q$CDM, and $\xi p$CDM), given various dataset combinations, using the \texttt{MCEvidence} code~\cite{Heavens:2017afc}. We then compute the natural logarithm of the Bayes factor with respect to $\Lambda$CDM, which we refer to as $\ln B$. With this definition, a value $\ln B>0$ [respectively $\ln B<0$] indicates that the interacting model is preferred [respectively disfavoured] over $\Lambda$CDM. We qualify the strength of the obtained values of $\ln B$ using the modified version of the Jeffreys scale provided in~\cite{Kass:1995loi}. In particular, the preference for the model with higher $\ln B$ is weak for $0 \leq \vert \ln B \vert <1$, positive for $1 \leq \vert \ln B \vert <3$, strong for $3 \leq \vert \ln B \vert <5$, and very strong for $\vert \ln B \vert \geq 5$.

\section{Results}
\label{sec:results}

We now discuss the results obtained using the methods and datasets described in Sec.~\ref{sec:data}. We begin by considering the baseline coupled vacuum $\xi\Lambda$CDM model, wherein the DE EoS is fixed to $w=-0.999$ (as a surrogate for the cosmological constant $\Lambda$ for which one has $w=-1$) and $\xi<0$. Then we will describe the $\xi q$CDM model, where $\xi<0$ and $w>-1$ , and finally the $\xi p$CDM model where $\xi>0$ and $w<-1$.

\subsection{Coupled vacuum: $\xi\Lambda$CDM model}
In this section we explore the same model as in Ref.~\cite{DiValentino:2019ffd} but in light of different datasets, notably including also the \textit{BAO} and \textit{Pantheon} measurements of the late-time expansion history. These results are summarized in Tab.~\ref{xi}.

Notice that with Planck CMB data alone, the value of the Hubble constant is much larger than that obtained in the absence of a DM-DE coupling ($H_0=67.27\pm 0.60)$~km/s/Mpc) and therefore the $H_0$ tension is strongly alleviated. When combining \emph{Planck} with R19 measurements, the statistical preference for a non-zero coupling $\xi$ is more significant than  $5\sigma$. These results agree with the ones obtained in~\cite{DiValentino:2019ffd}. The reason for this preference is given by the fact that in the coupled vacuum $\xi\Lambda$CDM model the energy flows from DM to DE, and then the amount of DM today is smaller. To match the position of the acoustic peaks in the CMB the quantity $\Omega_c h^2$ should not decrease dramatically, which automatically implies a larger value of $h$, i.e. $H_0$.

An important thing to point out is that $\Omega_ch^2$ is the physical density of cold DM \textit{today}. In the interacting models considered in this work, deviations from $\Lambda$CDM are almost exclusively occurring at late times, which is why the addition of late-time datasets such as \textit{BAO} or \textit{Pantheon} is important. As one can see from Eqs.~(\ref{eq:rhoc},\ref{eq:rhox}), for the region of parameter space considered, the cold DM energy density at the time of last-scattering in the interacting models is essentially the same as that in $\Lambda$CDM, explaining why these models are still able to fit the \textit{Planck} dataset well, as they leave the relative height of the acoustic peaks unchanged.

The addition of low-redshift measurements, as \textit{BAO} or Supernovae Ia Pantheon \textit{Pantheon} data, still hints to the presence of a coupling, albeit at a lower statistical significance. Also for these two data sets the Hubble constant values are larger than those obtained in the case of a pure $\Lambda$CDM scenario ($H_0= 67.66\pm0.42$~km/s/Mpc ($67.48\pm 0.50$~km/s/Mpc) for \textit{Planck}+\textit{BAO} (+\textit{Pantheon})). While in this case the central values of the inferred Hubble parameter are not as high as for the previously discussed case considering CMB data alone (for \textit{Planck}+\textit{BAO} we find $69.4^{+0.9}_{-1.5}$~km/s/Mpc), this value is large enough to bring the $H_0$ tension well below the $3\sigma$ level. In other words, the tension between \textit{Planck}+\textit{BAO} and R19 could be due to a statistical fluctuation in the case of an interacting scenario. Finally, when combining all datasets together (the \textit{All19} combination), we find $H_0=69.9 \pm 0.8$~km/s/Mpc, so that the tension with R19 is reduced to slightly more than $2.5\sigma$.

With regards to the \textit{BAO} dataset, it is important to remind the reader that BAO data is extracted under the assumption of $\Lambda$CDM, and the modified scenario of interacting dark energy could affect the result. However, the residual tension also clearly confirms earlier findings based on the inverse distance ladder approach (e.g.~\cite{Bernal:2016gxb,Feeney:2018mkj,Lemos:2018smw,Taubenberger:2019qna}) that finding late-time solutions to the $H_0$ tension which satisfactorily fit BAO and SNe data is challenging (albeit not impossible).

Finally, we compute $\ln B$ for all the 6 dataset combinations reported in Tab.~\ref{xi}. We confirm the findings of~\cite{DiValentino:2019ffd} that the preference for the coupled vacuum $\xi\Lambda$CDM model is positive when considering the \textit{Planck} dataset alone ($\ln B=1.3$), and very strong when considering the \textit{Planck}+\textit{R19} dataset combination ($\ln B=10.0$). The preference decreases to weak when considering the \textit{Planck}+\textit{lensing} dataset combination ($\ln B=0.9$). On the other hand, including late-time datasets through the \textit{Planck}+\textit{BAO} and \textit{Planck}+\textit{Pantheon} dataset combinations leads to the baseline $\Lambda$CDM model being preferred by Bayesian evidence considerations, with $\ln B=-0.6$ (weak preference) and $\ln B=-1.5$ (positive preference) respectively. Finally, considering the joint \textit{All19} dataset combination we find $\ln B=1.4$, and hence an overall positive preference for the $\xi\Lambda$CDM model. Although such a positive preference is mostly driven by the \textit{R19} dataset, we still find it intriguing given that the late-time \textit{BAO} and \textit{Pantheon} datasets (which strongly constrain late-time deviations from $\Lambda$CDM) were also included, and the resulting value of $H_0$ is such that the $H_0$ tension could be due to a statistical fluctuation in the case of the $\xi\Lambda$CDM model.

\squeezetable                                    
\begin{center}                              
\begin{table*}                                             
\begin{tabular}{cccccccccccccccc}       
\hline\hline                                                                                                                   
Parameters & Planck   & Planck & Planck& Planck & Planck & All19 \\ 
 &  &+R19 & +lensing  & +BAO & + Pantheon & \\ \hline
 
$\Omega_b h^2$ & $    0.0224 \pm 0.0002$ &  $    0.0224\pm0.0002$ & $    0.0224\pm 0.0002$ & $    0.0224\pm 0.0001$ & $    0.0224\pm 0.0002$ & $    0.0224\pm 0.0001$ \\
 
$\Omega_c h^2$ & $    <0.105 $  & $    0.031^{+0.013}_{-0.023}$ & $    <0.108$ & $    0.095^{+0.022}_{-0.008} $& $    0.103^{+0.013}_{-0.007} $ & $0.092^{+0.011}_{-0.009}$ \\

$\xi$ & $    -0.54^{+0.12}_{-0.28}$ &  $    -0.66^{+0.09}_{-0.13}$ & $    -0.51^{+0.12}_{-0.29}$ & $    -0.22^{+0.21}_{-0.05}$ & $    -0.15^{+0.12}_{-0.06}$ & $-0.24^{+0.09}_{-0.08}$ \\ \hline \hline

$H_0 $[km/s/Mpc] & $   72.8^{+3.0}_{+1.5}$&  $   74.0^{+1.2}_{-1.0}$ & $   72.8^{+3.0}_{+1.6}$ & $   69.4^{+0.9}_{-1.5}$ & $   68.6^{+0.8}_{-1.0}$ & $69.9 \pm 0.8$\\

$\sigma_8$ & $    2.27^{+0.40}_{-1.40}$ &  $    2.71^{+0.47}_{-1.30}$ & $    2.16^{+0.35}_{-1.40}$ & $    1.05^{+0.03}_{-0.24}$ & $    0.95^{+0.04}_{-0.12}$ & $1.04^{+0.08}_{-0.13}$\\

$S_8$ & $    1.30^{+0.17}_{-0.44}$ &  $    1.44^{+0.17}_{-0.34}$ & $  1.30^{+0.15}_{-0.42}  $ & $    0.93^{+0.03}_{-0.10}$ & $    0.89^{+0.03}_{-0.05}$ & $0.92^{+0.04}_{-0.06}$\\ \thickhline

$\ln B$ & $1.3$ & $10.0$ & $0.9$ & $-0.6$ & $-1.5$ & $1.4$ \\

Strength & Positive ($\xi\Lambda$CDM) & Very strong ($\xi\Lambda$CDM) & Weak ($\xi\Lambda$CDM) & Weak ($\Lambda$CDM) & Positive ($\Lambda$CDM) & Positive ($\xi\Lambda$CDM) \\

\hline\hline                                                  
\end{tabular}                                                   
\caption{Constraints on selected cosmological parameters of the $\xi\Lambda$CDM model. Constraints are reported as 68\%~CL intervals, unless they are quoted as upper/lower limits, in which case they represent 95\%~CL upper/lower limits. The horizontal lines separating the final three parameters ($H_0$, $\sigma_8$, and $S_8$) from the above ones highlight the fact that these three parameters are derived. The second-last row, separated from the above ones by a thicker line, reports $\ln B$, the natural logarithm of the Bayes factor computed with respect to $\Lambda$CDM for each of the datasets in question. A positive [respectively negative] value of $\ln B$ indicates that the $\xi\Lambda$CDM [respectively $\Lambda$CDM] model is preferred. The final row quantifies the strength of the preference for either the $\xi\Lambda$CDM model or the $\Lambda$CDM model (as appropriate given the sign of $\ln B$, and indicated in brackets) using the modified Jeffreys scale discussed in the text.}
\label{xi}                                              
\end{table*}                                                    
\end{center} 

\subsection{Coupled quintessence: $\xi q$CDM model}

The constraints on the \textit{quintessence} coupled model ($\xi q$CDM) are summarized in Tab.~\ref{wq}.

In these models, the energy flows from the DM to the DE sector and the amount of the DM mass-energy density today is considerably reduced as the values of the coupling $\xi$ are increased, see Eq.~(\ref{eq:rhoc}) and the left panel of Fig.~\ref{fig:tri}. This explains why the \textit{Planck}, \textit{Planck}+\textit{R19}, and \textit{Planck}+\textit{lensing} dataset combinations prefer a non-zero value of the coupling at a rather high significance level ($>3\sigma$), as a value $\xi<0$ can accommodate the smaller amount of DM required when $w>-1$. Also in this case, as for the $\xi\Lambda$CDM model, the cold DM energy density as last-scattering is essentially unchanged with respect to $\Lambda$CDM, which is why the model can fit \textit{Planck} data well.

Concerning the $H_0$ tension, even if the value of the Hubble constant $69.8^{+4.0}_{-2.5}$~km/s/Mpc obtained for Planck data only is larger than in the baseline $\Lambda$CDM model, it is still not as large as in the case of the $\xi\Lambda$CDM model discussed above. This is due to the strong anti-correlation between $w$ and $H_0$, see the left panel of Fig.~\ref{fig:wH0}. This well-known anti-correlation reflects the competing effects of $H_0$ and $w$ on the comoving distance to last-scattering and is dominating the impact of $\xi$, which would instead push $H_0$ to even larger values as we saw earlier.

When combining CMB with the low-redshift \textit{BAO} and \textit{Pantheon} datasets, intriguingly a significant preference for a large negative value of $\xi$ persists, contrarily to the  $\xi\Lambda$CDM scenario. Such a preference is driven by the fact that a non-zero coupling $\xi$ will reduce the large value of $\Omega_m$ required if the DE EoS is allowed to vary in the $w>-1$ region. As we saw earlier for the $\xi\Lambda$CDM model, adding low-redshift data decreases the central value of $H_0$, but it also reduces the significance of the Hubble tension between Planck+BAO and R19. Interestingly, we see that in case of \textit{Planck}+ \textit{BAO} and \textit{Planck}+\textit{Pantheon} there is also a preference for $w>-1$ at about three standard deviations. This preference is also suggested by the \textit{Planck}+\textit{R19} dataset. As a matter of fact, in the case of interacting dark energy, quintessence models agree with observations and also reduce the significance of the Hubble tension. When considering the \textit{All19} dataset combination, we find $H_0=69.8 \pm 0.8$~km/s/Mpc, and again as in the case of the $\xi\Lambda$CDM model the $H_0$ tension is reduced to slightly more than $2.5\sigma$.

Bayesian evidence considerations, however, overall disfavour the $\xi q$CDM model compared to $\Lambda$CDM. The extra parameter, $w$, is what is penalizing the $\xi q$CDM model. While the improvement in fit within the $\xi\Lambda$CDM model was sufficient to justify the extra parameter $\xi$, this is no longer the case in this model when taking into account the two extra parameters $\xi$ and $w$. In fact, except for the \textit{Planck}+\textit{R19} dataset combination, all other dataset combinations (including \textit{Planck} alone) favour $\Lambda$CDM, with strength ranging from weak (\textit{Planck} and \textit{All19}) to positive (\textit{Planck}+\textit{lensing}, \textit{Planck}+\textit{BAO}, and \textit{Planck}+\textit{Pantheon}), with the largest negative value of $\ln B$ being $\ln B=-2.6$ for the \textit{Planck}+\textit{Pantheon} dataset combination.

\squeezetable                                    
\begin{center}                              
\begin{table*}                                             
\begin{tabular}{cccccccccccccccc}       
\hline\hline                                                                                                                   
Parameters & Planck   & Planck & Planck& Planck & Planck & All19 \\ 
 &  &+R19 & +lensing  & +BAO & + Pantheon \\ \hline
 
$\Omega_b h^2$ & $    0.0224 \pm 0.0002$ &  $    0.0224\pm0.0002$ & $    0.0224\pm 0.0001$ & $    0.0224\pm 0.0001$ & $    0.0224\pm 0.0002$ & $0.0224 \pm 0.0001$ \\
 
$\Omega_c h^2$ & $    <0.099 $  & $    <0.045$ & $    <0.091$ & $    <0.099 $& $    <0.099 $ & $<0.087$\\

$\xi$ & $    -0.63^{+0.06}_{-0.22}$ &  $    -0.73^{+0.05}_{-0.10}$ & $    -0.61^{+0.08}_{-0.22}$ & $    -0.59^{+0.09}_{-0.25}$ & $    -0.58^{+0.10}_{-0.26}$ & $-0.59^{+0.10}_{-0.23}$\\

$w$ & $    <-0.69$ &  $    -0.95^{+0.01}_{-0.05}$ & $    <-0.71$ & $    -0.84^{+0.09}_{-0.07}$ & $    -0.84^{+0.09}_{-0.05}$ & $-0.87^{+0.08}_{-0.05}$\\ \hline \hline

$H_0 $[km/s/Mpc] & $   69.8^{+4.0}_{-2.5}$&  $   73.3^{+1.2}_{-1.0}$ & $   69.9^{+3.7}_{-2.5}$ & $   68.6\pm1.4$ & $   68.3\pm1.0$ & $69.8 \pm 0.7$\\

$\sigma_8$ & $    2.61^{+0.69}_{-1.70}$ &  $    3.43^{+0.94}_{-1.30}$ & $    2.48^{+0.63}_{-1.60}$ & $    2.31^{+0.56}_{-1.40}$ & $    2.21^{+0.46}_{-1.30}$ & $2.3^{+0.5}_{-1.3}$\\

$S_8$ & $    1.43^{+0.29}_{-0.46}$ &  $    1.63^{+0.31}_{-0.26}$ & $  1.39^{+0.23}_{-0.44}  $ & $    1.35^{+0.24}_{-0.45}$ & $    1.33^{+0.20}_{-0.44}$ & $1.34^{+0.19}_{-0.42}$\\ \thickhline

$\ln B$ & $-0.8$ & $7.4$ & $-1.3$ & $-1.8$ & $-2.6$ & $-0.3$ \\

Strength & Weak ($\Lambda$CDM) & Very strong ($\xi q$CDM) & Positive ($\Lambda$CDM) & Positive ($\Lambda$CDM) & Positive ($\Lambda$CDM) & Weak ($\Lambda$CDM) \\

\hline\hline                                                  
\end{tabular}                                                   
\caption{As in Tab.~\ref{xi}, for the $\xi q$CDM model.}
\label{wq}                                              
\end{table*}                                                    
\end{center} 

\subsection{Coupled phantom: $\xi p$CDM model}

The last model explored here is the one in which the DE EoS varies within the phantom region, $w<-1$. Therefore, to avoid instabilities, the coupling $\xi$ must be positive. The constraints on this model are shown in Tab.~\ref{wp}.

\begin{figure*}[th]
\begin{center}
	\includegraphics[width=0.49\textwidth]{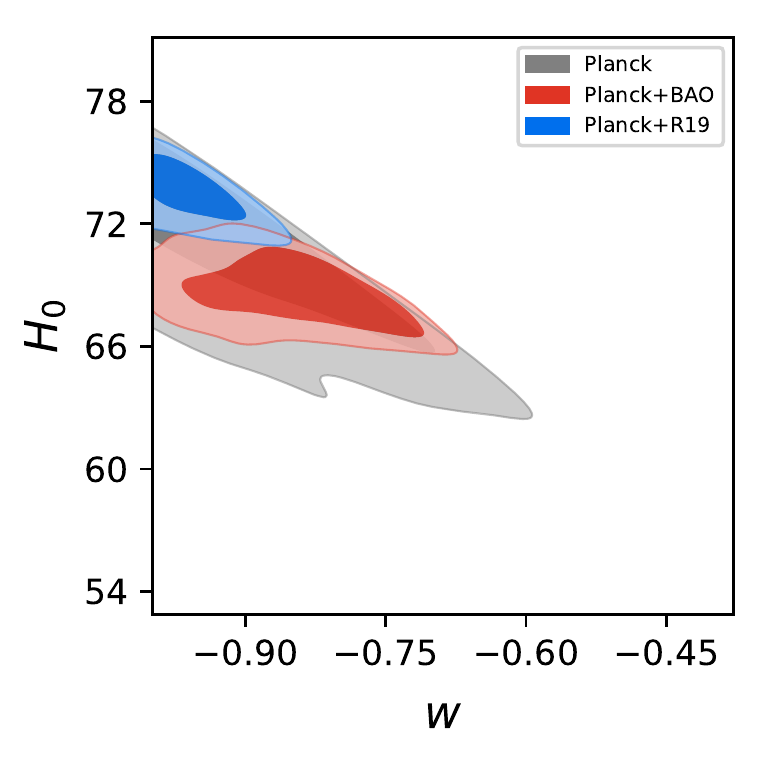}
	\includegraphics[width=0.49\textwidth]{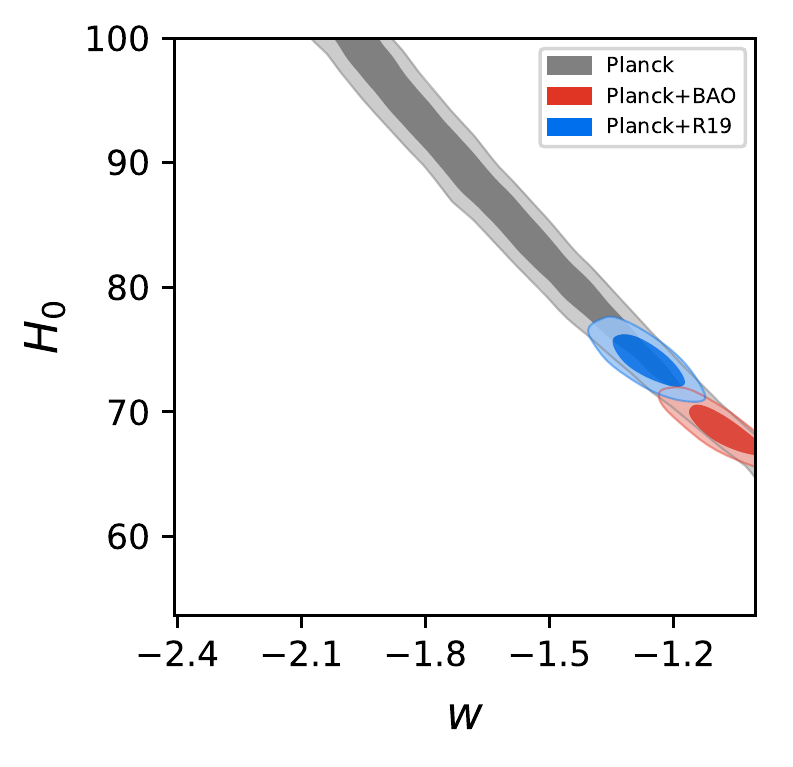}
	\caption{Left (right) panel: $68\%$ and $95\%$~CL allowed regions in the ($w, H_0$) plane for the $\xi q$CDM  ($\xi p$CDM) model For Planck alone, Planck+BAO, and Planck+R19. Note the marginal overlap between the
	Planck+BAO and Planck+R19 confidence regions indicating an easing of the Hubble tension.}
	\label{fig:wH0}
	\end{center}
\end{figure*}

Notice from the right panels of Fig.~\ref{fig:tri} and Fig.~\ref{fig:wH0} that \textit{(i)} the current amount of $\Omega_mh^2$ is slightly larger than within the $\Lambda$CDM case [see also Eq.~(\ref{eq:rhoc})]; and \textit{(ii)} the value of the Hubble constant is also always much larger than in the canonical $\Lambda$CDM. This is due to the well-known fact that when $w$ is allowed to vary in the phantom region, the parameter $H_0$ must be increased to not to affect the location of the CMB acoustic peaks. Consequently, we always obtain an upper bound on $\xi$ rather than a preferred region, as the presence of a non-zero coupling $\xi$ drives the value of $\Omega_m h^2$ to values even larger than those obtained when $w$ is not constant and is allowed to vary within the $w<-1$ region freely. Also in this case, as for the $\xi\Lambda$CDM and $\xi q$CDM models, the cold DM energy density as last-scattering is essentially unchanged with respect to $\Lambda$CDM, which is why the model can fit \textit{Planck} data well.

However, the $H_0$ tension is still also strongly alleviated in this case, as there is an extreme degeneracy between $w$ and $H_0$ (see the right panel of Fig.~\ref{fig:wH0}), with $H_0=81.3$~km/s/Mpc from Planck-only data. Therefore, as we saw earlier for the $\xi q$CDM model, the $H_0$-$w$ degeneracy is strongly dominating over the $H_0$-$\xi$ one. Therefore, within the $\xi p$CDM model, the resolution of the $H_0$ tension is coming from the phantom character of the DE component, rather than from the dark sector interaction itself.

When including low-redshift \textit{BAO} and \textit{Pantheon} measurements, the net effect is to bring the mean value of the DE EoS $w$ very close to $-1$. Consequently, the value of $H_0$ also gets closer to its standard mean value within the $\Lambda$CDM case, albeit remaining larger than the latter. In any case, we confirm that the $H_0$ tension is reduced with non-minimal dark energy physics also when low-redshift data are included. When considering the \textit{All19} dataset combination, we find $H_0=69.8 \pm 0.7$~km/s/Mpc, and again as in the case of the $\xi\Lambda$CDM and $\xi q$CDM models the $H_0$ tension is reduced to slightly more than $2.5\sigma$.

As we saw previously with the $\xi q$CDM model, Bayesian evidence considerations overall disfavour the $\xi p$CDM model compared to $\Lambda$CDM, even more so than they did for the $\xi q$CDM model. With the exception of the \textit{Planck}+\textit{R19} dataset combination, all other dataset combinations favour $\Lambda$CDM, with strength ranging from positive (\textit{Planck}, \textit{Planck}+\textit{lensing}, \textit{All19}), to strong (\textit{Planck}+\textit{BAO}), to very strong (\textit{Planck}+\textit{Pantheon}), with the largest negative value of $\ln B$ being $\ln B=-5.2$ for the \textit{Planck}+\textit{Pantheon} dataset combination.

For the sake of comparison, in In Tab.~\ref{lcdm} we report constraints on selected parameters of the three interacting dark energy models we have considered and compare them to the constraints instead obtained assuming $\Lambda$CDM. We do this only for the \textit{Planck} dataset.

Finally, using the full non-Gaussian posterior on $H_0$, we compute the tension with the local measurement of \textit{R19}, quoted in terms of number of $\sigma$s, for all possible combinations of the three interacting dark energy models and six dataset combinations studied in the paper. These numbers are reported in Tab.~\ref{tension}. As we see, the tension is at a level larger than $3\sigma$ only for the \textit{Planck}+\textit{Pantheon} dataset combination for all three models (even for the \textit{Planck}+\textit{BAO} dataset combination the tension always remains below the $2.9\sigma$ level). On the other hand, when considering the \textit{All19} dataset combination, the tension reaches at most the $2.7\sigma$ level, confirming our earlier claim that the residual tension in most cases could almost be justified by a statistical fluctuation.

\squeezetable                                    
\begin{center}                              
\begin{table*}                                             
\begin{tabular}{cccccccccccccccc}       
\hline\hline                                                                                                                   
Parameters & Planck   & Planck & Planck& Planck & Planck & All19 \\ 
 &  &+R19 & +lensing  & +BAO & + Pantheon \\ \hline
 
$\Omega_b h^2$ & $    0.0224 \pm 0.0002$ &  $    0.0224\pm0.0002$ & $    0.0224\pm 0.0002$ & $    0.0224\pm 0.0001$ & $    0.0224\pm 0.00012$ & $0.0224 \pm 0.0001$ \\
 
$\Omega_c h^2$ & $    0.132^{+0.005}_{-0.012} $  & $    0.133^{+0.006}_{-0.012}$ & $    0.133^{+0.006}_{-0.012}$ & $    0.134^{+0.007}_{-0.012} $& $    0.134^{+0.006}_{-0.012} $ & $0.132^{+0.006}_{-0.012}$ \\

$\xi$ & $    <0.248$ &  $    <0.277$ & $    <0.258$ & $    <0.295$ & $    <0.295$ & $<0.288$\\

$w$ & $    -1.59^{+0.18}_{-0.33}$ &  $    -1.26\pm0.06$ & $    -1.57^{+0.19}_{-0.32}$ & $    -1.10^{+0.07}_{-0.04}$ & $    -1.08^{+0.05}_{-0.04}$ & $-1.12^{+0.05}_{-0.04}$\\ \hline \hline

$H_0 $[km/s/Mpc] & $   >70.4$&  $   74.1\pm1.4$ & $   85.0^{+10.0}_{-5.0}$ & $   68.8^{+1.1}_{-1.5}$ & $   68.3\pm 1.0$ & $69.8 \pm 0.7$\\

$\sigma_8$ & $    0.88\pm0.08$ &  $    0.80^{+0.06}_{-0.04}$ & $    0.87\pm0.08$ & $   0.75\pm0.05$ & $    0.76^{+0.05}_{-0.04}$ & $0.76^{+0.06}_{-0.04}$\\

$S_8$ & $    0.74\pm 0.04$ &  $    0.78 \pm 0.03$ & $  0.74\pm 0.04  $ & $    0.79\pm 0.03$ & $    0.80\pm0.03$ & $0.79^{+0.03}_{-0.02}$\\ \thickhline

$\ln B$ & $-1.3$ & $5.6$ & $-1.6$ & $-4.5$ & $-5.2$ & $-2.7$ \\

Strength & Positive ($\Lambda$CDM) & Very strong ($\xi p$CDM) & Positive ($\Lambda$CDM) & Strong ($\Lambda$CDM) & Very strong ($\Lambda$CDM) & Positive ($\Lambda$CDM) \\

\hline\hline                                                  
\end{tabular}                                                   
\caption{As in Tab.~\ref{xi}, for the $\xi p$CDM model.}
\label{wp}                                              
\end{table*}                                                    
\end{center} 

\squeezetable                                    
\begin{center}                              
\begin{table*}                                             
\begin{tabular}{ccccccccccccccc}       
\hline\hline
Parameters & $\Lambda$CDM   & $\xi\Lambda$CDM & $\xi q$CDM & $\xi p$CDM \\ \hline
 
$\Omega_b h^2$ & $0.0224 \pm 0.0002$ &  $0.0224 \pm 0.0002$ & $0.0224 \pm 0.0002$ & $0.0224 \pm 0.0002$ \\
 
$\Omega_c h^2$ & $0.120 \pm 0.001$ & $<0.105$ & $<0.099$ & $0.132^{+0.005}_{-0.012} $ \\

$\xi$ & $0$ &  $-0.54^{+0.12}_{-0.28}$ & $-0.63^{+0.06}_{-0.22}$ & $<0.248$ \\

$w$ & $-1$ & $-0.999$ & $<-0.69$ & $-1.59^{+0.18}_{-0.33}$ \\ \hline \hline

$H_0 $[km/s/Mpc] & $67.3 \pm 0.6$& $72.8^{+3.0}_{-1.5}$ & $69.8^{+4.0}_{-2.5}$ & $>70.4$ \\

$\sigma_8$ & $0.81 \pm 0.01$ &  $2.27^{+0.40}_{-1.40}$ & $2.61^{+0.69}_{-1.70}$ & $0.88 \pm 0.08$ \\

$S_8$ & $0.83 \pm 0.02$ & $1.30^{+0.17}_{-0.44}$ & $1.43^{+0.29}_{-0.46}$ & $0.74 \pm 0.04$ \\

\hline\hline                                                  
\end{tabular}                                                   
\caption{Constraints on selected parameters of the $\Lambda$CDM, $\xi\Lambda$CDM, $\xi q$CDM, and $\xi p$CDM models, using the \textit{Planck} dataset alone. Constraints are reported as 68\%~CL intervals, unless they are quoted as upper/lower limits, in which case they represent 95\%~CL upper/lower limits.}
\label{lcdm}                                              
\end{table*}                                                    
\end{center}

\begin{center}                              
\begin{table*}                                             
\begin{tabular}{ccccccccccccccc}       
\hline\hline                                                                                                                   
Dataset & $\xi\Lambda$CDM & $\xi q$CDM & $\xi p$CDM \\ \hline
 
\textit{Planck} & $0.4\sigma$ &  $1.0\sigma$ & $0.5\sigma$ \\

\textit{Planck}+\textit{R19} & $<0.1\sigma$ &  $0.4\sigma$ & $<0.1\sigma$ \\

\textit{Planck}+\textit{lensing} & $0.4\sigma$ &  $1.0\sigma$ & $2.1\sigma$ \\

\textit{Planck}+\textit{BAO} & $2.7\sigma$ &  $2.7\sigma$ & $2.9\sigma$ \\

\textit{Planck}+\textit{Pantheon} & $3.3\sigma$ &  $3.3\sigma$ & $3.3\sigma$ \\

\textit{All19} & $2.5\sigma$ &  $2.7\sigma$ & $2.7\sigma$ \\

\hline\hline                                                  
\end{tabular}                                                   
\caption{Level of tension between the inferred value of $H_0$ and the \textit{R19} local measurements, quoted in terms of number of $\sigma$s, for all possible combinations of the three interacting dark energy models and six dataset combinations studied in the paper.}
\label{tension}                                              
\end{table*}                                                    
\end{center}

\section{Conclusions}
\label{sec:conclusions}

In this work, we have re-examined the hotly debated $H_0$ tension in light of the state-of-the-art high- and low-redshift cosmological datasets, within the context of extended dark energy models. In particular, we have considered interacting dark energy scenarios, featuring interactions between dark matter (DM) and dark energy (DE), allowing for more freedom in the dark energy sector compared to our earlier work~\cite{DiValentino:2019ffd}, by not restricting the dark energy equation of state to being that of a cosmological constant. Early-time superhorizon instability considerations impose stability conditions on the DM-DE coupling $\xi$ and the DE EoS $w$, which we have carefully taken into account.

The most important outcome of our studies is the fact that within these non-minimal DE cosmologies, the long-standing $H_0$ tension is alleviated to some extent. For most of the models and dataset combinations considered, we find indications for a non-zero DM-DE coupling, with a significance that varies depending on whether or not we include low-redshift BAO and SNeIa data. When we allow the DE EoS $w$ to change, we find that the $H_0$-$w$ degeneracy strongly dominates over the $H_0$-$\xi$ one. This implies that the $H_0$ tension is more efficiently solved in the coupled phantom $\xi p$CDM model with $\xi>0$ and $w<-1$ rather than in the coupled quintessence $\xi q$CDM model with $\xi<0$ and $w>-1$, due to the phantom character of the DE rather than due to the presence of the DM-DE interaction.

The inclusion of low-redshift BAO and SNe data (whose results the reader can find in the two rightmost columns of Tab.~\ref{xi}, Tab.~\ref{wq}, and Tab.~\ref{wp}) somewhat mildens all the previous findings, although it is worth remarking that the $H_0$ tension is still alleviated even in these cases. It is also intriguing to see that within the coupled quintessence $\xi q$CDM model with $\xi<0$ and $w>-1$, the indication for a non-zero DM-DE coupling persists even when low-redshift data is included. Interestingly, evidence for $w>-1$ at three standard deviations is present when BAO or SNeIa data are included.

Bayesian evidence considerations overall appear to disfavour the interacting models considered, although these conclusions depend very much on which of the three models and six dataset combinations one considers. For instance, the $\xi\Lambda$CDM model with 7 parameters appears to fare rather well when compared to $\Lambda$CDM, being favoured against $\Lambda$CDM for all dataset combinations except \textit{Planck}+\textit{BAO} and \textit{Planck}+\textit{Pantheon}. In particular, when combining all datasets together (the \textit{All19} combination), we find an overall positive preference for the $\xi\Lambda$CDM model over $\Lambda$CDM. The situation is much less favourable for the coupled quintessence and coupled phantom models with 8 parameters, which are always disfavoured (even rather strongly) against $\Lambda$CDM (the only exception being when considering the \textit{Planck}+\textit{R19} dataset combination). Overall, we conclude that the $\xi\Lambda$CDM model can still be considered an interesting solution to the $H_0$ tension \textit{even when low-redshift datasets and Bayesian evidence considerations are taken into account}. This is the main result of our paper.

As a word of caution, the full procedure which leads to the BAO constraints carried out by the different collaborations might be not necessarily valid in extended DE models such as the ones explored here. For instance, the BOSS collaboration, in Ref.~\cite{Anderson:2013zyy}, advises caution when using their BAO measurements (both the pre- and post-reconstruction measurements) in more exotic dark energy cosmologies (see also~\cite{Xu:2012hg} for related work exploring similar biases). Hence, BAO constraints themselves might need to be revised in a non-trivial manner when applied to constrain extended dark energy cosmologies. We plan to explore these and related issues in future work.

Overall, our results suggest that non-minimal modifications to the dark energy sector, such as those considered in our work, are still an intriguing route towards addressing the $H_0$ tension. As it is likely that such tension will persist in the near future, we believe that further investigations along this line are worthwhile and warranted.

\begin{acknowledgments}
E.D.V. acknowledges support from the European Research Council in the form of a Consolidator Grant with number 681431. A.M. is supported by TASP, iniziativa specifica INFN. O.M. is supported by the Spanish grants FPA2017-85985-P and SEV-2014-0398 of the MINECO and the European Union's Horizon 2020 research and innovation program under the grant agreements No.690575 and 67489. S.V. is supported by the Isaac Newton Trust and the Kavli Foundation through a Newton-Kavli fellowship, and acknowledges a College Research Associateship at Homerton College, University of Cambridge. This work is based on observations obtained with Planck (www.esa.int/Planck), an ESA science mission with instruments and contributions directly funded by ESA Member States, NASA, and Canada. We acknowledge use of the Planck Legacy Archive.
\end{acknowledgments}

\section*{Appendix}

Because there are strong correlations between certain parameters in all three interacting dark energy models studied, triangular plots showing the joint posteriors between these parameters might be more informative than the tables we presented. The most correlated parameters are $\Omega_ch^2$, $\xi$, $w$ (where applicable), as well as the derived parameters $H_0$ and $\Omega_m$. Here, we show triangular plots of the joint posteriors of these parameters within the three models studied, which clearly highlight the strong correlations at play.

\begin{figure*}[th]
\includegraphics[width=0.9\linewidth]{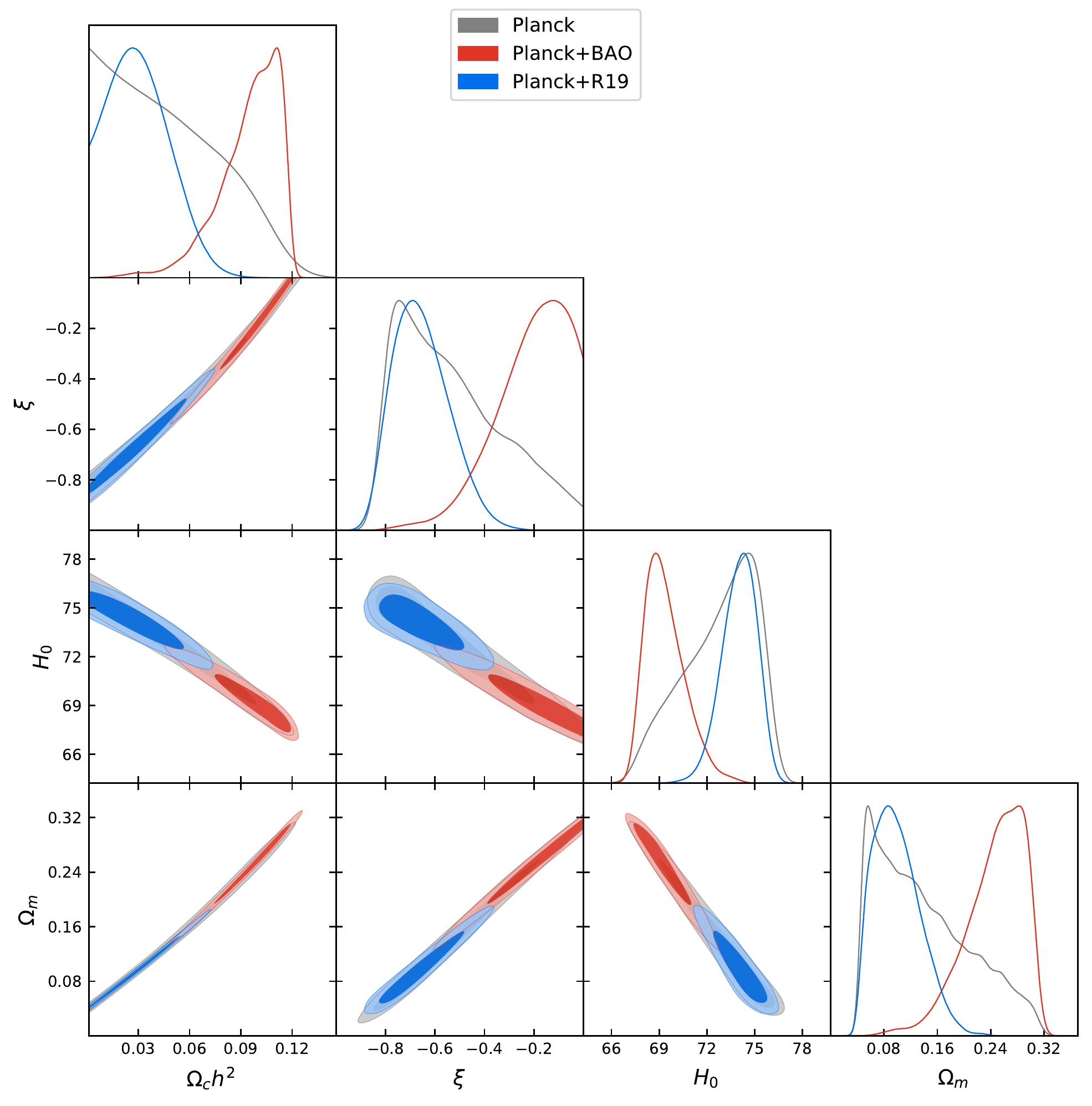}
\caption{Triangular plot showing the 2D joint and 1D marginalized posteriors of $\Omega_ch^2$, $\xi$, $H_0$, and $\Omega_m$, obtained assuming the coupled vacuum $\xi\Lambda$CDM model, for the \textit{Planck} (grey contours), \textit{Planck}+\textit{BAO} (red contours), and \textit{Planck}+\textit{R19} (blue contours) dataset combinations. The plot clearly highlights the strong correlations between these parameters.}
\label{fig:base_coupling_tri}
\end{figure*}

\begin{figure*}[th]
\includegraphics[width=0.9\linewidth]{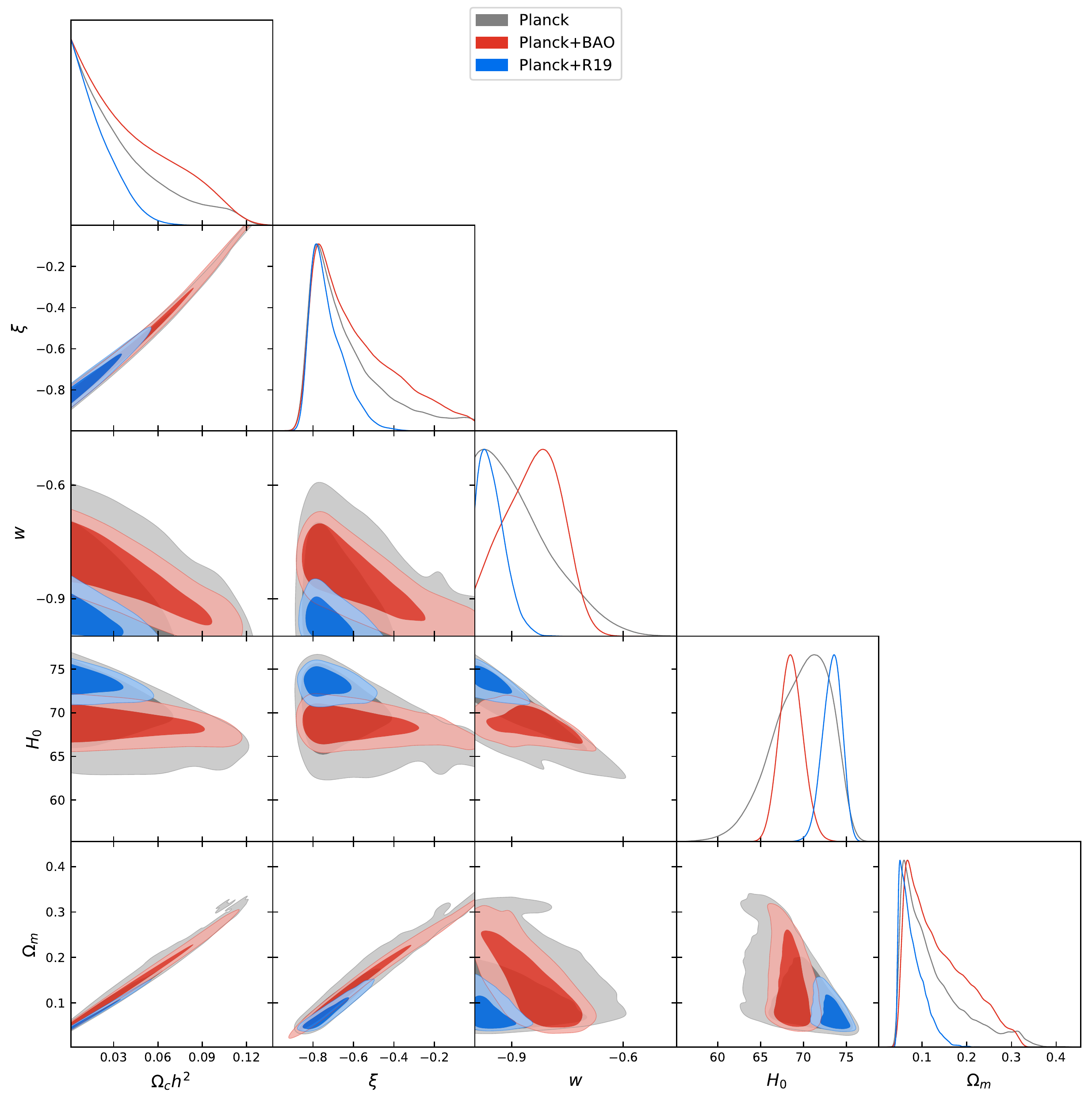}
\caption{Triangular plot showing the 2D joint and 1D marginalized posteriors of $\Omega_ch^2$, $\xi$, $w$ $H_0$, and $\Omega_m$, obtained assuming the coupled quintessence $\xi q$CDM model, for the \textit{Planck} (grey contours), \textit{Planck}+\textit{BAO} (red contours), and \textit{Planck}+\textit{R19} (blue contours) dataset combinations. The plot clearly highlights the strong correlations between these parameters.}
\label{fig:base_wpos_coupling_tri}
\end{figure*}

\begin{figure*}[th]
\includegraphics[width=0.9\linewidth]{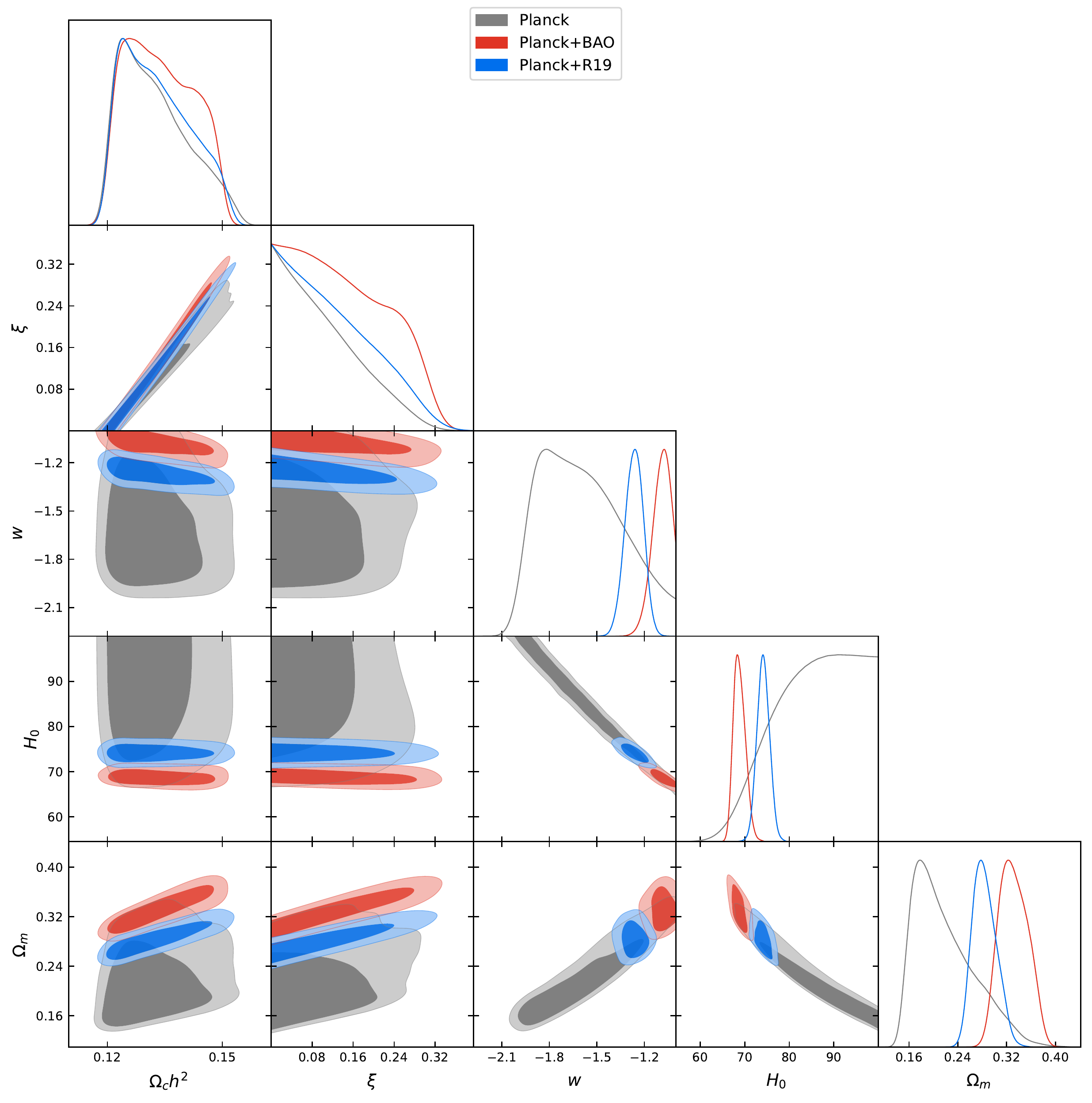}
\caption{Triangular plot showing the 2D joint and 1D marginalized posteriors of $\Omega_ch^2$, $\xi$, $w$, $H_0$, and $\Omega_m$, obtained assuming the coupled phantom $\xi p$CDM model, for the \textit{Planck} (grey contours), \textit{Planck}+\textit{BAO} (red contours), and \textit{Planck}+\textit{R19} (blue contours) dataset combinations. The plot clearly highlights the strong correlations between these parameters.}
\label{fig:base_wneg_coupling_tri}
\end{figure*}

\bibliographystyle{JHEP}
\bibliography{IDE.bib}

\end{document}